\documentclass[twocolumn,showpacs,floatfix,10pt,nofootinbib]{revtex4}

\def\bq{\begin{quote}}
\def\eq{\end{quote}}

 at 10truept

\newcommand{\beq}{\begin{equation}}
\newcommand{\eeq}{\end{equation}}
\newcommand{\beqa}{\begin{eqnarray}}
\newcommand{\eeqa}{\end{eqnarray}}

\usepackage{amssymb,amsmath}
\usepackage{graphicx,float}
\usepackage{indentfirst}
\newcommand{\be}{\begin{equation}}
\newcommand{\ee}{\end{equation}}

\newcommand{\ba}{\begin{eqnarray}}
\newcommand{\ea}{\end{eqnarray}}

\newcommand{\lie}{\pounds_{n}}

\begin{document}\title{Black holes in realistic branes: black string-like objects?}
\author{D. Bazeia}
\email{bazeia@fisica.ufpb.br}
\affiliation{Instituto de F\'\i sica, Universidade de S\~ao Paulo,  05314-970, S\~ao Paulo, SP, Brazil}
\affiliation{Departamento de F\'\i sica, Universidade Federal da
Para\'\i ba, 58051-970, Jo\~ao Pessoa, PB, Brazil}
\author{J. M. Hoff da Silva\footnote{Also at The Niels Bohr Institute - Copenhagen University, Blegdamsvej 17, DK-2100 Copenhagen, Denmark.}}
\email{hoff@feg.unesp.br}
\affiliation{Departamento de F\1sica e Qu\1mica, Universidade
Estadual Paulista, Av. Dr. Ariberto Pereira da Cunha, 333,
Guaratinguet\'a, SP, Brazil.}
\author{Rold\~ao da Rocha}
\email{roldao.rocha@ufabc.edu.br}
\affiliation{Centro de Matem\'atica, Computa\c c\~ao e Cogni\c c\~ao, Universidade Federal do ABC (UFABC) 09210-170, Santo Andr\'e, SP, Brazil.}

\begin{abstract}
A realistic model describing a black string-like object in an expanding Universe is analyzed in the context of the McVittie's solution of the Einstein field equations. The bulk metric near the brane is provided analogously to previous solutions for black strings. In particular, we show that at least when the Hubble parameter on the brane is positive, a black string-like object seems to play a fundamental role in the braneworld scenario,  generalizing the standard black strings in the context of a dynamical brane.

\end{abstract}\maketitle
\flushbottom


The search for solutions engendering realistic black holes on the brane, stable and presenting no naked singularity, 
is an object  of current  interest. Although an exact solution is known for a (1+2)-brane in a 4D bulk \cite{emp}, such task urges to be evinced in the 5D scenario with a single extra dimension of infinite extent. Numerical simulations of relativistic static stars on the brane and the exact analysis of the collapse on the brane as well --- based on the AdS/CFT correspondence \cite{emp2} --- appear as good efforts to address the issue \cite{maartens, yoshino}. There are arguments indicating that whichever the solutions are, they approach the Schwarzschild geometry at large distances \cite{maartens}.

On the other hand, black holes embedded in an expanding Universe can be described by 
 McVittie's solutions \cite{mcvittie}. In this cosmological scenario, a more realistic solution can be probed, providing an asymptotic Schwarzschild-de Sitter geometry on the brane. 
 The legitimate black hole interpretation holds at least when the cosmological scenario is dominated at late times by a positive cosmological constant \cite{Kaloper:2010ec}.  Nice features regarding the McVittie metric on the brane can be listed and potentially employed. For instance,  they reduce to the standard homogeneous and isotropic FRW cosmology, and to a Schwarzschild or de Sitter-Schwarzschild black hole, in appropriate limits. The issue is fascinating, endowing realistic models for black holes in the Universe. 
Interesting overviews on the subject are given, e.g., in \cite{Kaloper:2010ec,lake1,faraoni}, providing a deep and modern approach. 

In this Letter we are mainly concerned with the black string profile induced by the McVittie solution, by delving into the Taylor expansion outside a McVittie black hole metric on the brane along the extra dimension, where the corrections in the area of the associated 5D black string warped horizon arise. The issue induces interesting physical effects in the black string-like object, as we shall prove. 
The fine character of the expansion along the extra dimension is crucial to analyze the  generalized black string associated to 
the McVittie's solution on the brane. 
The way how the dynamical content of the solution 
on the brane affects the pathological properties regarding the black string \cite{Chamblin:1999by} shall be deeply reported.

In a braneworld with a single extra dimension of infinite extent, a vector field in the  bulk decomposes into components 
in the brane and orthogonal to the brane, as $ (x^{\alpha},y)$. The bulk is endowed with a
metric $\mathring{g}_{AB}dx^A dx^B = g_{\mu\nu}(x^\alpha,y)\,dx^\mu dx^\nu + dy^2$. The brane metric components $g_{\mu\nu}$ and the bulk metric are related by
$
\mathring{g}_{\mu\nu} = g_{\mu\nu} + n_\mu n_\nu, 
$ where $n^\sigma$ are the components related to a time-like vector field, splitting the bulk in normal coordinates, and  $g_{44} = 1$ and $g_{i4} = 0$.
In addition,  $\kappa^2_{4}=\frac{1}{6}\lambda\kappa^4_5$ and $
\Lambda_4=\frac{\kappa_5^{2}}{2}\Big(\Lambda_{5}+\frac{1}{6}\kappa_5^{2}\lambda^{2}\Big)$,
where $\Lambda_4$ is the effective brane cosmological constant,
$\kappa_4$ [$\kappa_5$] denotes the 4D [5D]
gravitational coupling, and $\lambda$ is the brane tension. Usually  $\kappa_5 = 8\pi G_5$, where 
$G_5$ denotes the 5D gravitational coupling, related to the
4D gravitational constant $G$ by $G_5 = G\ell_{\rm Planck}$ and $\ell_{\rm Planck} = \sqrt{G\hbar/c^3}$. The
extrinsic curvature is $K_{\mu\nu} = \frac{1}{2}\lie g_{\mu\nu}$
($\lie$ denotes  the Lie derivative, which in Gaussian
normal coordinates reads $ {\bf \pounds}_n=\partial/\partial
y$). The junction condition determines the extrinsic curvature on
the brane as
 \be\label{ext}
K_{\mu\nu}=-\frac{1}{2}\kappa_5^2 \left[T_{\mu\nu}+ \frac{1}{3}
\left(\lambda-T\right)g_{\mu\nu} \right].
 \ee 
 Here $T^{\mu\nu}$ is the energy-momentum tensor, and $T=T_\mu^{\;\mu}$. We also denote $K = K_\mu^{\;\mu}$ and $K^2 = K_{\alpha\beta}K^{\alpha\beta}$.
Given the 5D Weyl tensor
$
 C_{\mu\nu\sigma\rho} = {}^{(5)}R_{\mu\nu\sigma\rho} - \frac{2}{3} (\mathring{g}_{[\mu\sigma} {}^{(5)}R_{\nu]\rho} + \mathring{g}_{[\nu\rho} {}^{(5)}R_{\mu]\sigma}) - \frac{1}{6} {}^{(5)}R ( \mathring{g}_{\mu[\sigma} \mathring{g}_{\nu\rho]})$ where ${}^{(5)}R_{\mu\nu\sigma\rho}$ denotes the components of the 5D Riemann tensor (${}^{(5)}R_{\mu\nu}$ and ${}^{(5)}R$ are the associated Ricci tensor and the scalar curvature), the symmetric and trace-free components   are respectively 
 the electric (${\cal E}_{\mu\nu} = C_{\mu\nu\sigma\rho} n^\sigma n^\rho$) and magnetic (${\cal B}_{\mu\nu\alpha} = g_\mu^{\;\rho} g_\nu^{\;\sigma}
C_{\rho\sigma\alpha\beta}n^\beta$) Weyl tensor components.  
The effective field equations are complemented by a set of equations, obtained from the 5D Einstein equations and Bianchi equations \cite{GCGR,Gergely:2006hd, maartens}, which are employed to calculate the terms of the Taylor expansion of the metric along the extra dimension, providing in particular the black string profile and some physical consequences, given by (hereon we denote $g_{\mu\nu}(x^\alpha,0) = g_{\mu\nu}$):
\begin{widetext}
 \ba
\hspace*{-0.4cm}{\;\;\;\;}g_{\mu\nu}(x^\alpha,y)&\!\!=\!\!& g_{\mu\nu}-\kappa_5^2\left[
T_{\mu\nu}\!+\!\frac{1}{3}(\lambda-T)g_{\mu\nu}\right]|y| \!+\! \left[\frac{1}{4}\kappa_5^4\!\left(
T_{\mu\alpha}T^\alpha_\nu -{\cal E}_{\mu\nu} +\frac{2}{3} (\lambda-T)T_{\mu\nu}
\right)\! +\!\left( \frac{1}{36}
\kappa_5^4(\lambda-T)^2-\frac{\Lambda_5}{6}
\right)\!g_{\mu\nu}\right] y^2 {\;}\nonumber\\
&& +\left.\Bigg[2K_{\mu\beta}K^{\beta}_{\;\,\alpha}K^{\alpha}_{\;\,\nu} - ({\cal E}_{\mu\alpha}K^{\alpha}_{\;\,\nu}+K_{\mu\alpha}{\cal E}^{\alpha}_{\;\,\nu})-\frac{1}{3}\Lambda_5K_{\mu\nu}-\nabla^\alpha{\cal B}_{\alpha(\mu\nu)} + \frac{1}{6}
\Lambda_5\left(K_{\mu\nu}-g_{\mu\nu}K\right)+K^{\alpha\beta}R_{\mu\alpha\nu\beta}
 \right.\nonumber\\
&&\left.\qquad\qquad + 3K^\alpha{}_{(\mu}{\cal
E}_{\nu)\alpha}-K{\cal E}_{\mu\nu}+\left(K_{\mu\alpha}K_{\nu\beta}
-K_{\alpha\beta}K_{\mu\nu}\right)K^{\alpha\beta}-\frac{\Lambda_5}{3}K_{\mu\nu}\Bigg]\;\frac{|y|^3}{3!} + \cdots \right.  \label{tay} \ea
\end{widetext}
Such expansion regards the metric on the bulk near the brane, and was analyzed in \cite{casadio1, maartens} only up to the second order. The fourth order expansion was derived in \cite{maartens} for a particular case. In \cite{meuhoff} the most complete fourth order expansion, also containing the additional terms coming from the variable brane tension, was accomplished. Due to the awkward 
expression therein, we insert above the expansion up to the third order. 
 As a particular case, the black hole horizon evolution along the extra dimension (the warped horizon \cite{clark}) may be examined, by exploring the component $g_{\theta\theta}(x^\alpha,y)$ in (\ref{tay}).
Indeed, any spherically symmetric metric associated to a black hole presents radial coordinate given by $\sqrt{g_{\theta\theta}(x,0)} = {\rm r}$. 
The black hole solution, namely, the black string solution \emph{on the brane}, is regarded when 
$\sqrt{g_{\theta\theta}(x,0)} = R$, where $R$ denotes the coordinate singularity. More precisely, the black string
 horizon for the Schwarzschild metric  is defined when  ${\rm r} =  \frac{2GM}{c^2}$,  obtained when the 
  coefficient $\left(1-\frac{2GM}{c^2{\rm r}}\right) = g_{{\rm rr}}$ of the term $d{\rm r}^2$ goes to infinity \cite{Casadioharms}. It corresponds to the black hole horizon on the brane. On the another hand, the radial coordinate $r$ in spherical coordinates legitimately appears as the term $g_{\theta\theta}d\theta^2 = {\rm r}^2 d\theta^2$ in the Schwarzschild metric. 
 Our analysis of the term $g_{\theta\theta}(x^\alpha,y)$ by Eq.~(\ref{tay})  holds for any value $r$ and provides the bulk metric. In particular, the term originally coined ``black string'' 
 corresponds to the Schwarzschild metric on the brane \cite{clark}, defined by the black hole horizon evolution along the extra dimension into the bulk. Hence, the black string  regards solely the  so called ``warped horizon'', which is $g_{\theta\theta}(x^\alpha,y)$, for the particular case where ${\rm r} = R$ is a coordinate singularity. 
 
 Now we argue whether such interpretation regarding black strings
 holds for the McVittie's solutions.  In their simplest form they have zero spatial curvature in the asymptotically FRW region, but  can be generalized \cite{mcvittie, kastra}. The spatial curvature of the FRW geometry is not expected to appreciably alter the behavior of the metric near a mass source as long as the gravitational radius of the mass $M$ whichever larger, be smaller than the radius of curvature.
The metric is written in isotropic spherical coordinates \cite{buchdahl} defined by $ {\rm r} = r\left(1+ \frac{GM}{{r}}\right)$, as \cite{mcvittie, Kaloper:2010ec}
\begin{equation} ds^2 = - \Bigl(\frac{1-\mu}{1+\mu}\Bigr)^2 dt^2 + (1+\mu)^4
a^2(t) (dr^2 +r^2d\Omega^2), \label{mv} \end{equation}
where $a(t)$ is the asymptotic cosmological scale factor, $\mu = \frac{M}{2a(t) r}$, and  
$M$ is the mass parameter of the source. Using spatial translations, $r = 0$ is chosen as the center of spherical symmetry. 
Here the asymptotically spatially flat FRW metric is considered, suggesting a cosmic scenario compatible to current cosmological data \cite{car,wil}.
It is an exact solution of the field equations for an arbitrary mass $M$ provided that $a(t)$ solves the
Friedmann equation and
\be \rho(t) = \frac{3\dot{a}^2}{8 \pi Ga^2}   \, , \label{friedman} \ee
which describes the matter energy density, with $H = \frac{\dot a}{a}$ being the Hubble parameter.  The isotropic pressure associated to the fluid can be written as \cite{sakaihaines}
\be
p = -\frac{1}{8\pi G} \left(3 \frac{\dot{a}^2}{a^2} +2 \frac{1+\mu}{1-\mu} \left(\frac{\ddot{a}}{a}-\frac{\dot{a}^2}{a^2}\right)  \right), 
\label{pressure}
\ee
having a homogeneous term proportional to $ H^2$ and an inhomogeneous term as well, 
containing $ \dot H$. 
The McVittie's solution has a curvature singularity at $\mu = 1$, since the Ricci scalar can be expressed in the form $R = 12H^{2}+6  \dot H\left(\frac{1+\mu}{1-\mu}\right)$; this singularity is interpreted as a cosmological big bang singularity \cite{Kaloper:2010ec}. 

McVittie's solution  describes black holes embedded in expanding FRW Universes, when the Hubble parameter is positive.
Some results advocate  
spherically symmetric solutions in asymptotically FRW cosmologies \cite{nolann}. McVittie's solution is one sample among the geometries describing masses in FRW, where the mass parameter is a constant {and} the energy density is homogeneous. The inhomogeneous pressure is hence necessary.  
 The initial Big-Bang singularity is absent when $\dot H =0$, and in fact the geometries (\ref{mv}) reduce to either the Schwarzschild or Schwarzschild-de Sitter solutions. In the case $a(t)=1$ the McVittie's solution reduces to a black hole in flat space, and the metric (\ref{mv}) provides the Schwarzschild solution in isotropic coordinates. 

A black string-like object associated to the McVittie's solution is led into the Schwarzschild and FRW ones as limiting cases.
Therefore, we adopt an effective approach, studying the horizon variation Taylor expansion. 
The Weyl term on the brane is given by \cite{maartens}
\begin{eqnarray}
{\cal E}_{\theta\theta}(r,t)= -\left[\rho^2\left(\frac{1}{6}\left(\frac{1+\mu}{1-\mu}\right)^2\left(2- \frac{1+\mu}{1-\mu}\right)^2\right)\right.\nonumber\\\left.+\frac{1}{4(1+\mu)^4a^2}\right]- \frac{\rho p}{4a^2}\left(1+4\mu+5\mu^2+4\mu^3\right)
\label{weyl}
\end{eqnarray} for the McVittie's solution (\ref{mv}). By substituting the expressions (\ref{friedman}, \ref{pressure}) for $\rho(t)$ and $p(t)$ above, as the black string horizon variation along the extra dimension is
analyzed, the term $g_{\theta\theta}(x^\alpha,y)$ in (\ref{tay}) is given by 
\begin{widetext}\ba
g_{\theta\theta}({\rm r},t,y)\!&=&\!{\rm r}^2\left[1-\kappa_5^2\left(3\frac{\dot{a}^2}{a^2}+\frac{\lambda}{3}+\frac{1+\mu}{1-\mu}\left(\frac{\ddot{a}}{a}-\frac{\dot{a}^2}{a^2}\right)+\frac{3\dot{a}^2}{2}\left(\frac{1+\mu}{1-\mu}\right)^6\right)\right.\,|y|\nonumber\\
&+&\left[\frac{3\dot{a}^4}{16 a^4}\!\left[\frac{1}{6}\left(\frac{1+\mu}{1-\mu}\right)^2\!\left(2\!-\! \frac{1+\mu}{1-\mu}\right)^2\!+\!\frac{1}{4(1+\mu)^4a^2}\!+\! \left(3\frac{\dot{a}^2}{a^2}\!+\!\frac{2(1+\mu)}{1-\mu}\!\left(\frac{\ddot{a}}{a}\!-\!\frac{\dot{a}^2}{a^2}\right)\!\right)\!
\left(\frac{3+4\mu+5\mu^2+4\mu^3}{(1+\mu)^5(1-\mu)}\right)\right.\right.\nonumber\\
&+&\left.\left.\left(\frac{27}{4(1+\mu)^4a^2}-\frac{3(1-\mu)}{2(1+\mu)}+\frac{9(1+\mu)^2}{4(1-\mu)^2}\right) \right]+ \frac{3(1-\mu)}{2(1+\mu)^5a^2}+ \frac{2}{3}\lambda- \frac{\Lambda_5}{6}(1+\mu)^4 a^4\right.\nonumber\\
&+&\left.\left.\frac{1}{36}\!\kappa_5^4\!(1+\mu)^4\!\left(\lambda + \frac{3\dot{a}^2}{2}\frac{(1+\mu)^2}{(1-\mu)^2} + \frac{9\dot{a}^2}{2(1+\mu)^4} +  \frac{3}{(1-\mu)(1+\mu)^3}\left(\frac{\ddot{a}}{a}-\frac{\dot{a}^2}{a^2}\right)\right)^2 \right]\,\frac{y^2}{2!} + \cdots\right]
\label{expp}
 \ea
 \end{widetext}
 The brane metric $g_{\mu\nu}$ is regular everywhere on and outside the black hole horizon and away from the big bang. The expansion including the term $y^4$ was considered in \cite{meuhoff}, and here Eq.(\ref{expp}) is written up to the second order in the extra dimension for the sake of conciseness.  
When  $a(t)=1$, by transforming the spherical isotropic coordinates to the spherical standard ones, the bulk metric component in (\ref{expp})  --- which is the warped horizon of the black string-like object when {\rm r} is constant --- is led to the bulk metric component 
 \[
g_{\theta\theta}({\rm r},y) = {\rm r}^2\!\left(1-\!\frac{\lambda}{3}\kappa_5^2\,|y|+\!\left(\!\frac{1}{36}\kappa_5^4\lambda^2 - \frac{1}{6}\Lambda_5\!\right)y^2 + \cdots\right),\] which, in particular, is the classical Schwarzschild black string warped horizon when {\rm r} is constant. Therefore the results in \cite{Chamblin:1999by} --- further generalized in \cite{Chamblin:2001} and discussed in, e. g.,  \cite{maartens, clark, meuhoff} --- are obtained, in such limit.

Our aim is to analyze the possibility and the properties of a black string-like object locally associated to the McVittie's solution, in the context of (\ref{tay}) and (\ref{expp}).
In the case of a pure FRW metric, the state parameter $w = \frac{p}{\rho}$ is defined and the Einstein field equations on the brane provide $
\rho \propto a^{-3(1+w)/\beta}$ (where we define $\beta := \frac{1-\mu}{1+\mu}$) leading together with the Friedmann equation to the time evolution of the scale factor. When the mass $M=0$ it implies $\beta = 1$, and the scale factor takes the well known value for the scale factor of a flat universe $a(t)\propto t^{{2}/{3}}$ (dominated by non-relativistic matter, where $w=0$) or $a(t)\propto t^{{1}/{2}}$ (dominated by the radiation or relativistic matter, where $w=\frac{1}{3}$). In the case of a cosmological constant ($w=-1$), we have $a(t)\propto \exp(H_0 t)$, independently of $\beta$. The manifold $\mu=1$ $(\beta=0)$, corresponds to the event horizon in the Schwarzschild case. However, we should avoid this value in order to circumvent the big bang singularity \cite{Kaloper:2010ec}. In all cases below, the value $\beta = 0.9$ is used to illustrate the results. Different values for $\beta$ do not affect the physical aspects underlying the results below.

We shall compare the black string-like object profile in the two eras of evolution of our Universe, without a cosmological constant and, in addition, in the presence of a cosmological constant.
Let us first consider the case where the scale factor $a(t) \propto t^{\beta/2}$, emulating a radiation-dominated brane. In the figures below the time parameter $t$ is considered in the scale [$t$] = $10^8\; yr$,  corresponding to our choice $\Lambda = 1 =\kappa_5$. The black string-like object has the warped horizon provided below, depicted in Fig.~1 for a fixed time $t=0.8$ and different values of $\ell$ . Further, the graphic depicted in Fig.~2 evinces the warped horizon  profile  according to the time evolution, along the extra dimension.

Next, we consider the scale factor as $a(t) \propto t^{2\beta/3}$, mimicking a non-relativistic matter-dominated brane. The results are illustrated in Figs. 3 and 4. Also, in Figs. 5 and 6 we depict the results for the scale factor in the form $a(t) \propto \exp(H_0 t)$, which leads to a braneworld scenario dominated by a cosmological constant.

The question regarding a black string-like object, correspondent to the McVittie's solution, can be more  realistically answered, by considering 
the legitimate black hole interpretation of the McVittie's solution \cite{Kaloper:2010ec}. It occurs when the Hubble parameter is positive. In the graphics below, the value for the $g_{\theta\theta}$ component is provided in units of $2GM$ and $\Lambda = 1=\kappa_5$.

\begin{figure}[H]\begin{center}
\includegraphics[width=2.65in]{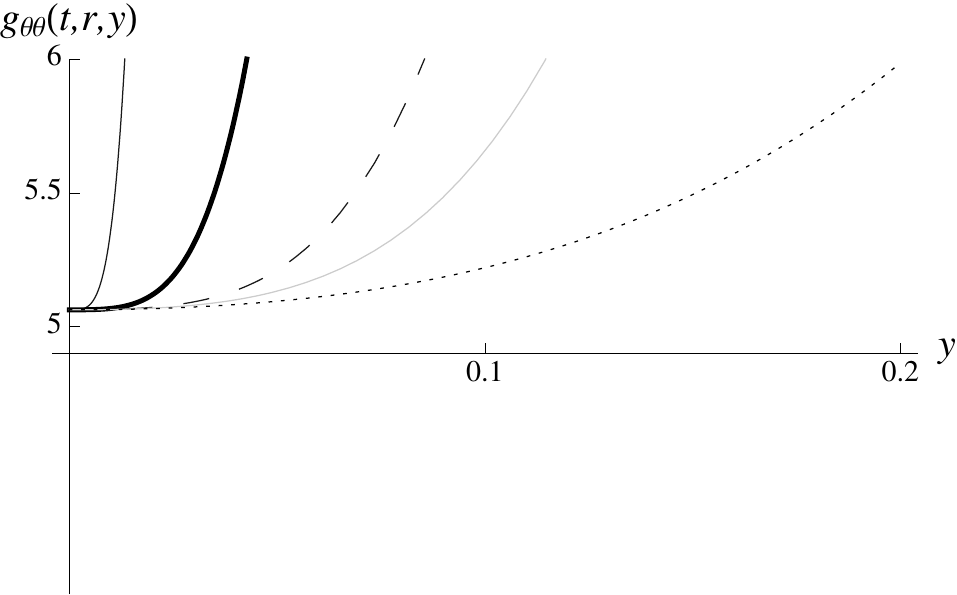}
\caption{\footnotesize\; Plots of the warped horizon $g_{\theta\theta}(t,r,y)$ for ($a(t) \propto t^{\beta/2}$) along the extra dimension $y$, for different values of the bulk curvature radius parameter $\ell$. For the dotted line, $\ell = 10^{-2}$ mm; for the light-gray line, $\ell = 10^{-3}$ mm; for the dashed black line,  $\ell=10^{-4}$ mm;  for the black line,  $\ell=10^{-5}$ mm; and for the dark-gray line $\ell= 10^{-6}$ mm.}\end{center}\end{figure}
\begin{figure}[H]\begin{center}
\includegraphics[width=2.45in]{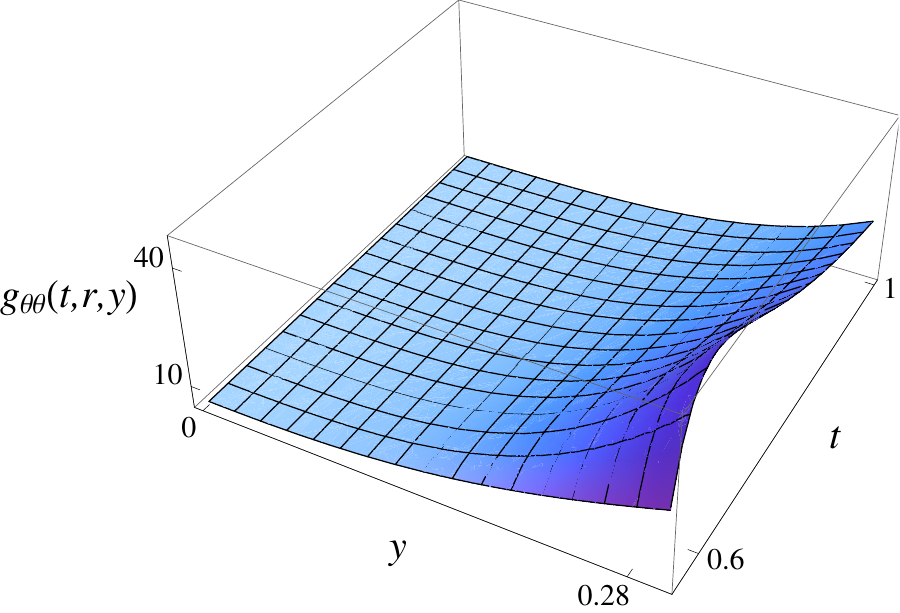}
\caption{\footnotesize\; Plot of the warped horizon $g_{\theta\theta}(t,r,y)$ along the extra dimension $y$, as an explicit function of time $t$,  for $a(t) \propto t^{\beta/2}$.}
\end{center}\end{figure}

The original  black string, corresponding to the Schwarzschild singularity, is obtained in the limit where 
$a(t) = 1$.
The formation of a galaxy with a central black hole was investigated in \cite{hell}. The black hole  may be either a collapsed object or a generalization of a wormhole \cite{hell}; the cases wherein such interpretations hold are depicted in Figs. 1-4. The black string-like object warped horizon can be accomplished by the Eqs.~(\ref{tay}, \ref{expp}), given a black hole 
in the brane and its associated singularity. 

Figs. 1 and 3 evince the time-dependent profile that mimics a radiation-dominated and a matter-dominated scenario, respectively. {They depict the graphics for  the bulk metric $g_{\theta\theta}(t,r,y)$ along the extra dimension, for different values of the bulk curvature radius parameter $\ell$ for a fixed time $t=0.8$, in full compliance with the upper limit of $\ell$ in the region $\ell\lesssim  0.2$ mm \cite{8888}. Figs. 2 and 4 clearly reinforces  such profiles, which correspond respectively to Figs. 1 and 3, which are solely their  section for $t=0.8$. It can be noticed that in a non-relativistic matter-dominated brane (Fig. 2) the warped horizon of the black string-like object increases more abruptly then in a radiation-dominated brane (Fig. 1).}
Fig. 5 shows the time-dependent profile that describes a pure cosmological constant scenario on the brane. The warped horizon evolutes along the extra dimension even more smoothly then in the radiation-dominated era, due to the presence of a cosmological constant. In all cases, the evolution drastically contributes for the warped horizons alterations, as the time elapses, along the extra dimension.

\begin{figure}[H]\begin{center}
\includegraphics[width=2.65in]{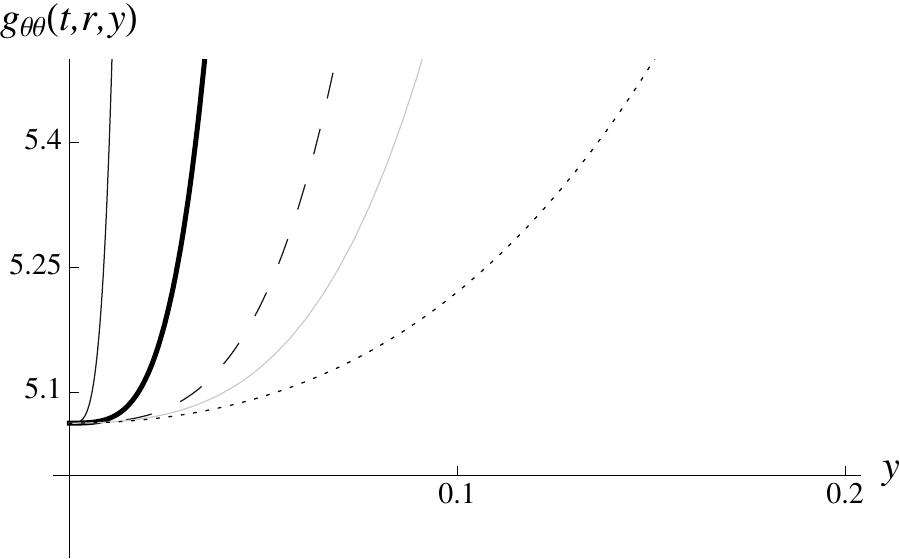}
\caption{\footnotesize\; Plots of the  warped horizon $g_{\theta\theta}(t,r,y)$ for $(a(t) \propto t^{2\beta/3})$ along the extra dimension $y$, for different values of the bulk curvature radius parameter $\ell$. For the dotted line, $\ell = 10^{-2}$ mm; for the light-gray line, $\ell = 10^{-3}$ mm; for the dashed black line, $\ell=10^{-4}$ mm;  for the black line, $\ell=10^{-5}$ mm; and for the dark-gray line, $\ell= 10^{-6}$ mm.}
\includegraphics[width=2.45in]{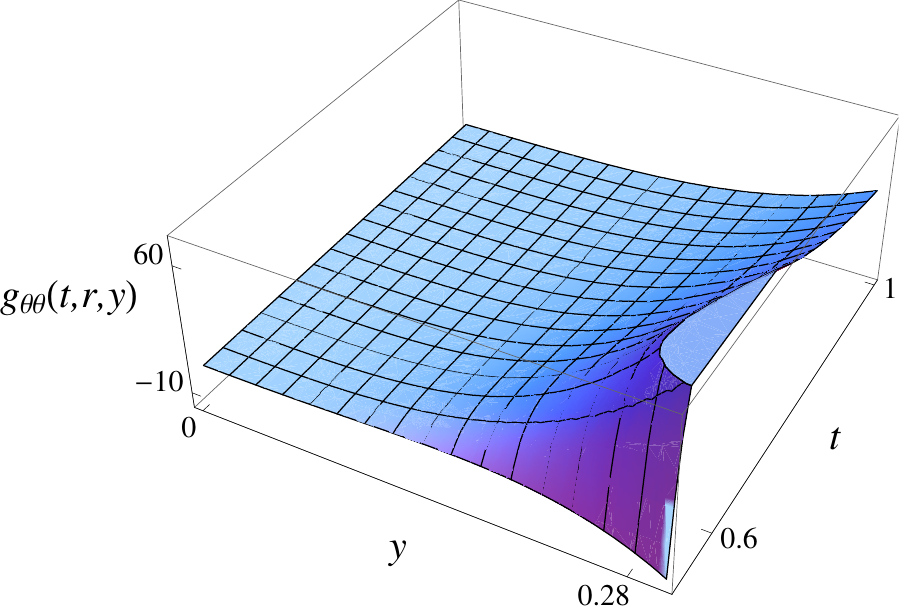}
\caption{\footnotesize\; Plot of the warped horizon $g_{\theta\theta}(t,r,y)$ along the extra dimension $y$, as an explicit function of time $t$, for $a(t) \propto t^{2\beta/3}$.}
\end{center}\end{figure}
{In Figs. 2 and 4, for each slice of constant time in the range considered in the graphics, there is a subtle and prominent difference between  both the  warped horizons, regarding the respective corresponding eras. In the Fig. 2, regarding a brane dominated by the radiation or relativistic matter $(a(t) \propto t^{2\beta/3})$, the warped horizon of the associated black string-like object increases monotonically along the extra dimension, irrespectively of the time. Instead,  
in the Fig. 4, which concerns a brane dominated by non-relativistic matter $(a(t) \propto t^{\beta/2})$, as the time evolutes the warped horizon of the associated black string-like object decreases along the extra dimension for any value for $t<0.54$. For the time parameter greater than this value, the warped horizon of the black string-like object always increases, along the extra dimension.  }

The pure cosmological constant braneworld scenario exhibited in Figs. 5 and 6
approaches a realistic black string-like object in a global asymptotically FRW braneworld,
where locally the behavior of a solution is analyzed. From the  Einstein equations, by providing on the brane a relation analogous to the Friedmann equation, depending  only on the geometry and matter content of the brane, the solution on the brane is extended to the bulk in \cite{bhr}, where exact solutions to the brane cosmology are compatible  with standard cosmology in a Randall--Sundrum--like  braneworld scenario, with a single extra dimension of infinite extent. Our results engender such achievements, without the necessity of taking into account the extra dimensional dependence on the scale factor nonetheless.
As the metric at the bulk  provides coefficients for the terms $|y|^k$ in Eqs.~(\ref{tay}, \ref{expp}), such extra dimensional dependence in the scale factor does not add relevant  physical information to the current results.

\begin{figure}[H]\begin{center}
\includegraphics[width=2.5in]{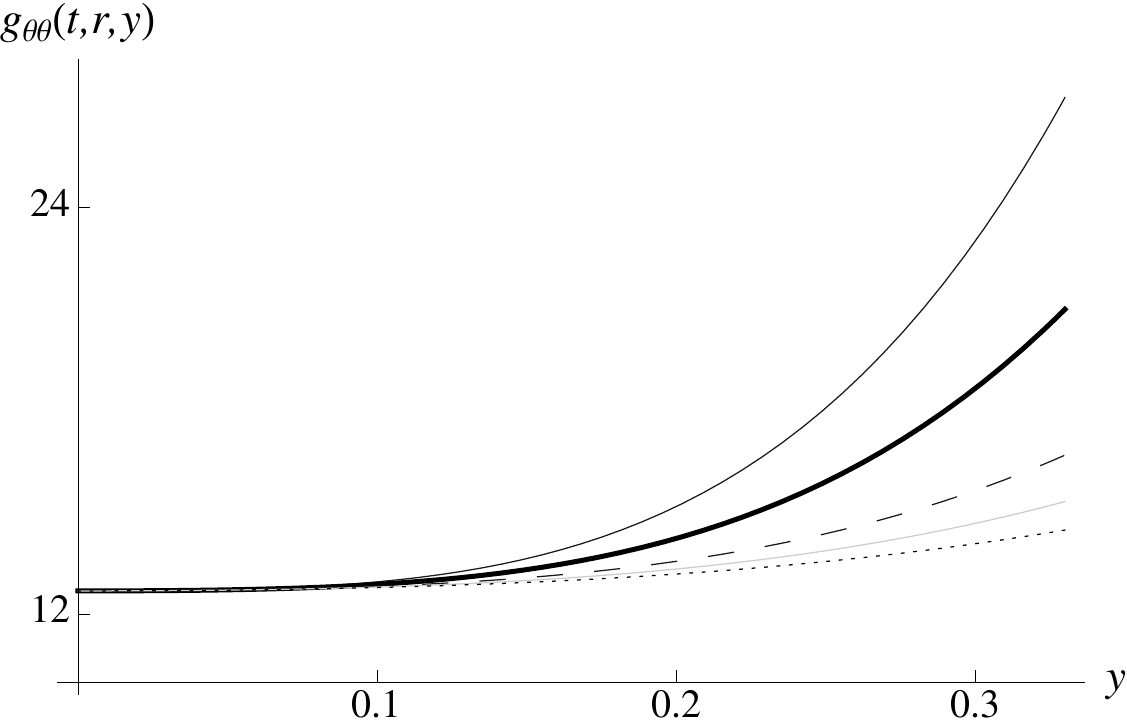}
\caption{\footnotesize\; Plots of the warped horizon $g_{\theta\theta}(t,r,y)$ in a scenario dominated by a cosmological constant, for $a(t) \propto \exp(H_0t)$, along the extra dimension $y$. For the dotted line, $\ell = 10^{-2}$ mm; for the light-gray line, $\ell = 10^{-3}$ mm; for the dashed black line,  $\ell=10^{-4}$ mm;  for the black line,  $\ell=10^{-5}$ mm; for the dark-gray line, $\ell= 10^{-6}$ mm.}
\includegraphics[width=2.8in]{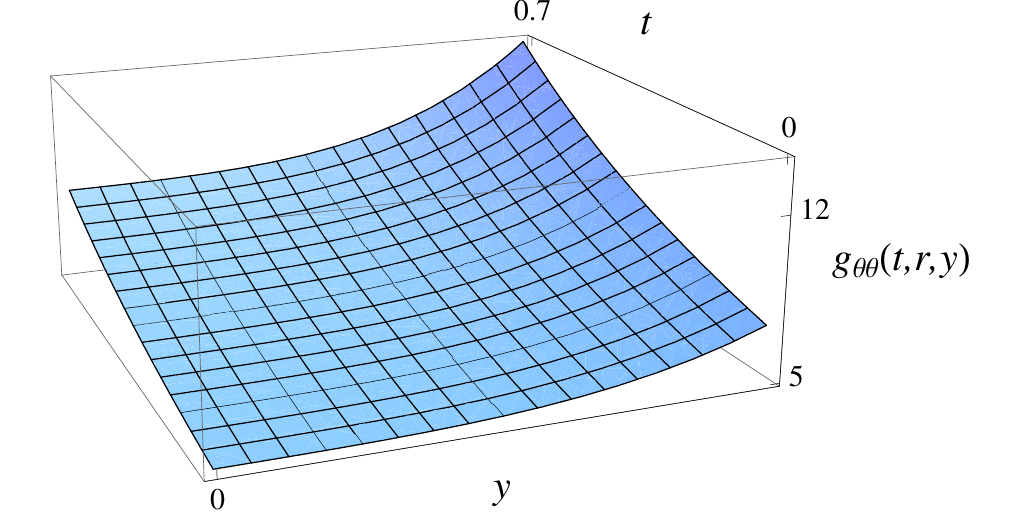}
\caption{\footnotesize\; Plot of the warped horizon $g_{\theta\theta}(t,r,y)$ along the extra dimension $y$, as an explicit function of time $t$. Here $a(t) \propto \exp(H_0t)$, as in the previous figure.}
\end{center}\end{figure}

In summary, as the McVittie solution is established to present a legitimate interpretation as a black hole when $H_0 > 0$ \cite{Kaloper:2010ec}, we have shown that there is indeed a realistic black string-like object in this case. Moreover, our 
method describes the metric in the bulk when the brane evinces radiation-dominated and matter-dominated scenarios as well, suggesting the possibility for similar interpretations.

 In order to reveal the physical nature of the black string-like object introduced above, one cannot rely on a non-invariant quantity such as the metric component to deduce conclusions on the topology
of a spacetime. Rather, the use of an invariant quantity  would be a better choice, as the one analyzed in \cite{Kaloper:2010ec}. For completeness, we have checked that when the scale factor $a(t)=1$, the associated Kretschmann scalars are led to the Schwarzschild ones. 
In fact, when at late times the cosmology is dominated by a positive cosmological constant, the McVittie metric on the brane is regular everywhere on and outside the black hole horizon, and it asymptotes in the future and near the horizon to the Schwarzschild-de Sitter geometry, which has a black string-like object associated,  already thoroughly accomplished and analyzed in \cite{EPJC}. It essentially holds for any slice  of the extra dimension, in gaussian coordinates. In the Schwarzschild-de Sitter geometry, there can be a point $y_1$ along the extra dimension for which the Kretschmann scalar ${}^{(5)}K = {}^{(5)}R_{\mu\nu\rho\sigma} {}^{(5)}R^{\mu\nu\rho\sigma}$ diverges at $r\rightarrow 0$, and also at $y=y_1$, irrespective of the value for $r$, characterizing a  singularity. Furthermore, for the case where the cosmological constant equals zero, our results are in full compliance with \cite{Chamblin:1999by}. In fact, in such particular case  the Kretschmann scalar  ${}^{(5)}K \propto {48G^2M^2
\over r^6}\, e^{4|y|/\ell}$ diverges at the bulk horizon as well as at the black string singularity at $r =0$ \cite{Chamblin:1999by}.

Here, when $M=0$, the solution obtained reduces to a standard homogeneous and isotropic FRW cosmology on the brane, and for $H(t)$ constant, it is led to a Schwarzschild (standard) black string \cite{maartens, Chamblin:1999by, meuhoff} or de Sitter-Schwarzschild (or Kottler) black string of mass $M$, already investigated in  \cite{EPJC}. All curvature invariants on the null surface  equal their values on the horizon of a Schwarzschild-de Sitter generalized black string of mass $M$ \cite{EPJC} and positive Hubble constant, and  this null surface is a soft, null naked singularity in an FRW spacetime if  $H(t)= 0$ at late times. At least in the case  when $H_{0}>0$ the McVittie metric on the brane induces a black string-like object.

Now we can further probe the bulk properties, as an alternative radial coordinate is defined \cite{nolann} as
${\tt r} = (1+\mu)^2 a(t) r$. It makes the McVittie metric (\ref{mv}) to read
\begin{equation}
d s^2 = -g d t^2 - {2H{\tt r}}f^{-1/2}d {\tt r}\,d t + f^{-1}{d {\tt r}^2}  + {\tt r}^2 d\Omega_2,
\label{fin}\nonumber
\end{equation}\noindent where 
$f = 1-2M/{\tt r}$. 
On the brane, a null apparent horizon is placed at ${\tt r}_-$, the smaller positive root of $g({\tt r})=1- 2M/{\tt r} - H^2 {\tt r}^2=0$. 
 When $H$ equals a constant, the metric above is the Schwarzschild-de Sitter metric in coordinates similar the Eddington-Finkelstein ones
 In the case $H_{0}>0$ [$H_{0}=0$] it is a regular black hole event horizon [null singularity].  
By using this scaling one sees that the Ricci scalar $R=12 H^{2}+6 \dot H f^{-1/2}$ is finite in the limit ${\tt r}=2M$, but higher curvature invariants may be not finite. For instance, the invariant  \be \xi=(\nabla_{\mu} \nabla_{\nu} R_{\tau\psi\rho \sigma})(\nabla^{\mu} \nabla^{\nu} R^{\tau\psi\rho \sigma})\label{xi4}\ee contains a term  $H^{4} {\dot{H}}^{2} f^{-5}$ which diverges at the horizon along ingoing null geodesics \cite{Kaloper:2010ec}. When $H_{0}>0$ the invariant (\ref{xi4}) attains a maximum at a finite distance from the horizon ${\tt r}={\tt r}_{-}$ and then decreases to its Schwarzschild-de Sitter value.  When $H_{0}\to 0$, the horizon becomes a null,  soft, and weak singularity \cite{Kaloper:2010ec}.
The 4D and the 5D Riemann tensors are related by the Gauss equation as
 \begin{equation}
{}^{(5)}R_{\tau\psi\rho\sigma} = R_{\tau\psi\rho\sigma}
 -K_{\tau\rho}K_{\psi\sigma} + K_{\tau\sigma}K_{\psi\rho},\label{gauss}\end{equation} and
 the 5D version of the invariant $\xi$ in (\ref{xi4}) 
 reads
 \beq
 {}^{(5)}\xi=(D_a D_b {}^{(5)}R_{\tau\psi\rho \sigma})(D^aD^b {}^{(5)}R^{\tau\psi\rho\sigma})\label{xi5}\eeq where the indices $a,b$ are effectively  4D spacetime indexes, since the decomposition of the 5D covariant derivative can be expressed as $D_a = \nabla_\mu$, for $a = 0,\ldots, 3$, and $D_a = \nabla_y$, when $a=4$. Therefore, the difference between the invariants in (\ref{xi4}) and (\ref{xi5}) is given by \begin{widetext}
 \begin{eqnarray}
{}^{(5)}\xi - \xi\!&=&\!
 2(\nabla_{\mu} \nabla_{\nu} K_{\tau\rho}K_{\psi\sigma})(\nabla^{\mu} \nabla^{\nu} K^{\tau\rho}K^{\psi\sigma})\!- \!2(\nabla_{\mu} \nabla_{\nu} K_{\tau\sigma}K_{\psi\rho})(\nabla^{\mu} \nabla^{\nu}K^{\tau\rho}K^{\psi\sigma}) +2(\nabla_{\mu} \nabla_{\nu} K_{\tau\rho}K_{\psi\sigma})(\nabla^{\mu} \nabla^{\nu}K^{\tau\rho}K^{\psi\sigma}) \nonumber\\
&-& 2(\nabla_{\mu} \nabla_{\nu} K_{\tau\sigma}K_{\psi\rho})(\nabla^{\mu} \nabla^{\nu}R^{\tau\psi\rho\sigma}) 
-
 4(\nabla_y \nabla_{\nu} K_{\tau\rho}K_{\psi\sigma})(\nabla^y \nabla^{\nu} R^{\tau\psi\rho \sigma})
-2 (\nabla_{y} \nabla_{\nu} K_{\tau\rho}K_{\psi\sigma})
(\nabla^{y} \nabla^{\nu}K^{\tau\sigma}K^{\psi\rho})
\nonumber\\&+&2(\nabla_y \nabla_\nu  K_{\tau\sigma}K_{\psi\rho})(\nabla^y \nabla^\nu R^{\tau\psi\rho \sigma})
+(\nabla_y \nabla_{\nu} R_{\tau\psi\rho \sigma})(\nabla^y \nabla^{\nu} R^{\tau\psi\rho \sigma})-4(\nabla_\mu \nabla_{y} K_{\tau\rho}K_{\psi\sigma})(\nabla^\mu \nabla^{y} R^{\tau\psi\rho \sigma})
\nonumber\\&-& 
2 (\nabla_{\mu} \nabla_{y} K_{\tau\rho}K_{\psi\sigma})
(\nabla^{\mu} \nabla^{y}K^{\tau\sigma}K^{\psi\rho})+(\nabla_\mu \nabla_{y} R_{\tau\psi\rho \sigma})(\nabla^\mu \nabla^y R^{\tau\psi\rho \sigma})+ 2(\nabla_{\mu} \nabla_{y}K_{\tau\sigma}K_{\psi\rho})
(\nabla^{\mu} \nabla^{y}K^{\tau\sigma}K^{\psi\rho})
\nonumber
 \\&+&
 (\nabla_y^2 R_{\tau\psi\rho \sigma})((\nabla^y)^2 R^{\tau\psi\rho \sigma})-4(\nabla_y^2 K_{\tau\rho}K_{\psi\sigma})((\nabla^y)^2 R^{\tau\psi\rho \sigma})+2(\nabla_y^2 K_{\tau\sigma}K_{\psi\rho})((\nabla^y)^2 K^{\tau\sigma}K^{\psi\rho})\nonumber\\&-&2(\nabla_y^2 K_{\tau\rho}K_{\psi\sigma})((\nabla^y)^2 K^{\tau\sigma}K^{\psi\rho})\label{xi6}\end{eqnarray} \end{widetext} By considering the extrinsic curvature in (\ref{ext}), the terms on the right hand side in (\ref{xi5}) do not cancel the divergence of the 4D invariant in (\ref{xi4}), hence the black string invariant ${}^{(5)}\xi$ diverges at the black string warped horizon as well as in the black string-like singularity, in full compliance with its limiting case when $a(t)=1$, which originates the classical black string \cite{Chamblin:1999by}.

Our analysis goes beyond  the existence of black string-like objects in realistic braneworlds, since Eqs.~(\ref{tay}, \ref{expp}) describe the metric in the bulk.
Wherever the radial coordinate on the brane describes a black hole  horizon, hence the black string-like warped horizon is obtained as a particular situation.
\section*{Acknowledgements}
D. Bazeia would like to thank CAPES, CNPq and FAPESP for financial support. R. da Rocha is grateful to CNPq (476580/2010-2 and 304862/2009-6) for financial support. J. M. Hoff da Silva is thankful to CNPq (482043/2011-3 and 308623/2012-6) for partial support.

\end{document}